\PassOptionsToPackage{svgnames, prologue, dvipsnames}{xcolor}
\documentclass{vgtc}                          
 
\graphicspath{{figures/}{pictures/}{images/}{./}} 

\usepackage{times}                     

\usepackage{tabu}                      
\usepackage{booktabs}                  
\usepackage{lipsum}                    
\usepackage{mwe}            

\usepackage{csquotes}
\usepackage{amsmath}
\usepackage{bm}
\usepackage{fontawesome}
\usepackage{graphicx}
\usepackage{glossaries}
\usepackage{multirow}
\usepackage{mathtools} 

\usepackage{mathptmx}                  

\usepackage{pgfplots}    

\usepackage{xcolor}
\DeclareUnicodeCharacter{2212}{−}
\usepgfplotslibrary{groupplots,dateplot}
\usetikzlibrary{shapes,shapes.misc, graphs, patterns,shapes.arrows, arrows.meta, positioning, fit, arrows.meta, graphs, shapes.misc, matrix, quotes, patterns, shapes.arrows, shapes.misc, positioning, calc, fit, bending}

\newacronym[shortplural=LMs, longplural=Language Models]{lm}{LM}{Language Model}
\newacronym[shortplural=KGs, longplural=Knowledge Graphs]{kg}{KG}{Knowledge Graph}
\newacronym[shortplural=GUIs, longplural=Graphical User Interfaces]{gui}{GUI}{Graphical User Interface}
\newacronym{ged}{GED}{Graph Edit Distance}
\newacronym{rag}{RAG}{Retrieval Augmented Generation}
\newacronym{oql}{OQL}{Ontology Query Language}
\newacronym[shortplural=BGPs, longplural=Basic Graph Patterns]{bgp}{BGP}{Basic Graph Pattern}
\newacronym{gbnf}{GBNF}{GGML Backus-Naur Form}
\newacronym{onset}{OnSET}{Ontology and Semantic Exploration Toolkit}
\newacronym{bto}{BTO}{Brainteaser Ontology}
\newacronym{als}{ALS}{Amyotrophic lateral sclerosis}
\newacronym{ms}{MS}{Multiple sclerosis}

\onlineid{0}

\vgtccategory{Research}

\vgtcinsertpkg

\title{Difference Views for Visual Graph Query Building}

\author{%
   \authororcid{Benedikt Kantz}{0000-0003-3294-8421},
   \authororcid{Stefan Lengauer}{0000-0001-5136-4320},
   \authororcid{Peter Waldert}{0009-0004-8459-7381},
  and 
   \authororcid{Tobias Schreck}{0000-0003-0778-8665}
}
\affiliation{\scriptsize Graz University of Technology, Austria}

\teaser{
  \centering
 \pgfdeclarelayer{background}
\pgfdeclarelayer{foreground}
\pgfsetlayers{background,main,foreground}
\newcommand{\blockcolor}{CadetBlue}

\rmfamily
\usetikzlibrary{shapes.geometric, arrows.meta, positioning, fit, shapes.symbols}
\begin{tikzpicture}[
  mapping/.style={rectangle, draw=\blockcolor!90,  fill=\blockcolor!5, minimum size=1cm,  text width=2.4cm, align=center, font={\small},  line width=3pt, minimum height=1.75cm},
  example/.style={mapping, text width=1.9cm, minimum height=0.4cm},
  >={Stealth[round]},
  every new ->/.style={shorten >=1pt, line width=1pt},
  every new --/.style={line width=1pt},
  gray_dotted/.style={
      draw=black!30!white, line width=1pt,
      dash pattern=on 1pt off 4pt on 6pt off 4pt,
      inner sep=2mm, rectangle
    },
  ]
  \node (intermediate) [fill=CornflowerBlue!5, draw=CornflowerBlue!90] {
    \newcommand{\erlsize}{0.5}
    \input{figures/erl_edit.tex}
  };
  \draw (intermediate.south)+(0,-0.1) coordinate (intermediate_spacer);
  \node (intermediate_lbl) at (intermediate.north) [above, inner sep=1mm, text width=3.25cm, align=center]{
    Difference Graph $(a)$
  };
  \begin{pgfonlayer}{background}
    \node (intermediate_box) [mapping, fit=(intermediate_lbl) (intermediate) (intermediate_spacer)] {};
  \end{pgfonlayer};
  \draw (intermediate_box.north)+(-2,0.25) coordinate  (intermediate_tr);
  \draw (intermediate_box.north)+(+2,0.25) coordinate  (intermediate_tl);
  \draw (intermediate_box.north-|intermediate_tr)  coordinate  (intermediate_tr_touch);
  \draw (intermediate_box.north-|intermediate_tl)  coordinate  (intermediate_tl_touch);

  \draw (intermediate_box.south)+(-2.2,0) coordinate  (intermediate_bl);
  \draw (intermediate_box.south)+(+2.2,0) coordinate  (intermediate_br);
  \node(lm_assist)[mapping, below=0.15cm of intermediate_box.south,text width=3cm, minimum height=0.7cm] {\small Structured LM output \& semantic retrieval $(b)$~\faDatabase};
  \draw (lm_assist.west)+(0,+0.2) coordinate  (lm_assist_node_top);
  \draw (lm_assist.west)+(0,-0.2) coordinate  (lm_assist_node_bot);
  \node(query)[mapping, left=of lm_assist_node_bot,draw=BurlyWood!90,fill=BurlyWood!10, text width=1.8cm,  minimum height=0.5cm] {\small NL Request~\faLanguage};

  \draw (lm_assist_node_top-|intermediate_bl) coordinate  (intermediate_bl_con);
  \draw (lm_assist-|intermediate_br) coordinate  (intermediate_br_con);

  \node (ex_1) [example, right=1.2cm of intermediate] {
    \newcommand{\erlsize}{0.15}
    \newcommand{\erlperson}{P1}
    \newcommand{\erlcompany}{C1}
    \newcommand{\erlplace}{P1}
    \newcommand{\erlschool}{U1}
    \input{figures/erl_ex.tex}
  };
  \node (ex_2) [example, above=0.1cm of ex_1] {
    \newcommand{\erlsize}{0.15}
    \newcommand{\erlperson}{P2}
    \newcommand{\erlcompany}{C2}
    \newcommand{\erlplace}{P2}
    \newcommand{\erlschool}{U2}
    \input{figures/erl_ex.tex}
  };
  \node (ex_3) [example, below=0.1cm of ex_1] {
    \newcommand{\erlsize}{0.15}
    \newcommand{\erlperson}{P3}
    \newcommand{\erlcompany}{C3}
    \newcommand{\erlplace}{P3}
    \newcommand{\erlschool}{U3}
    \input{figures/erl_ex.tex}
  }; 
  \draw ($(intermediate.east)!0.5!(ex_1.west)$)  coordinate  (between_ex_1);
  \draw (ex_2.west-|between_ex_1)  coordinate  (between_ex_2);
  \draw (ex_3.west-|between_ex_1)  coordinate  (between_ex_3);
  \node (desc_ex) [gray_dotted,
    fit = (ex_1) (ex_2) (ex_3) ] {};
  \draw (desc_ex.north east)+(4,0) coordinate (result_box_tr);
  \draw (desc_ex.south east)+(4,0) coordinate (result_box_br);
  \node (result_text) at (desc_ex.south east) [below, inner sep=2mm] {Result Views $(c)$};
  \draw ($(desc_ex.north east)!0.5!(result_box_br)$) coordinate (result_box_center);
  \node (result_box) at (result_box_center) [rectangle, centered] {\input{figures/school_colours.tex}};
  \draw (desc_ex.north east) [gray_dotted]  -- (result_box_tr) -- (result_box_br) -- (desc_ex.south east);
  \graph [use existing nodes] {
  intermediate_bl--intermediate_bl_con->lm_assist_node_top;
  lm_assist--intermediate_br_con->intermediate_br;
  query->lm_assist_node_bot;
  intermediate_tr_touch--intermediate_tr--["manual refinements", sloped, anchor=top, inner sep=2pt]intermediate_tl->intermediate_tl_touch;
  intermediate.east--between_ex_1->ex_1.west;
  between_ex_1--{between_ex_2->ex_2.west, between_ex_3->ex_3.west};
  };
\end{tikzpicture}
  \caption{Overview of the graph editing pipeline. The user can modify an existing query and keep track of the changes made using a difference graph with additions highlighted in green, and deletions in red compared to prior query and the current state $(a)$, which can be modified by a LM assistant using constrained outputs $(b)$, yielding comparative results views that consider the difference graph for multiple view types $(c)$.}
  \label{fig:teaser}
}

\abstract{
  \glspl{kg} contain vast amounts of linked resources that encode knowledge in various domains, which can be queried and searched for using specialized languages like SPARQL, a query language developed to query \glspl{kg}.  Existing visual query builders enable non-expert users to construct SPARQL queries and utilize the knowledge contained in these graphs. Query building, however, is an iterative and often visual process where the user's question can change and differ throughout the process, especially for exploratory search. Our visual querying interface communicates these changes between iterative steps in the query-building process using graph differences to contrast the changes and the evolution in the graph query. We also enable users to formulate their evolving information needs using a natural language interface directly integrated into the difference query view. We, furthermore, communicate the change in results in the result view by contrasting the differences in both result distribution and individual instances of the prototype graph and demonstrate the system's applicability through case studies on different ontologies and usage scenarios, illustrating how our system fosters, both, data exploration and analysis of domain-specific graphs.
} 


\begin{document}


\firstsection{Introduction}
\label{sec:introduction}
\maketitle

\glspl{kg} contain large amounts of interlinked knowledge from different domains that is usually difficult to explore without dedicated query systems. Structured \glspl{kg} can be constrained by ontologies, enabling a structured querying approach~\cite{hogan2021knowledge}. These can cover a variety of domains ~\cite{Lehmann2015DBpediaA, UniProt2025, Faggioli2024}, mapping real-world entities and links into structured knowledge that can be used for a broad range of applications, like question-answering, pattern search to industrial applications~\cite{Abu-Aisheh2015}. These graphs can be queried using SPARQL, a specialized query language developed for these \glspl{kg}~\cite{hogan2021knowledge, Seaborne2024Sparql}. This query language, however, poses a challenge for non-expert users searching for information in these graphs. Visual query builders address these issues by enabling non-expert users to create queries using standard user interfaces. Some visual query builders even display users' graph results directly in their interfaces \cite{Vargas2019RDF} and provide guidance on initial queries and throughout the query building process \cite{Kantz2025}.
 
These discussed visual query builders, however, do not consider the \emph{changes} and \emph{evolution} of queries between different stages of information need, or differences in the result set. These attributes are commonly found in \emph{exploratory search}, where the user seeks information driven by curiosity. The formulation of queries is often iteratively performed based on the results obtained from a starting query, as the final information goal may not be clearly defined at the beginning of the search~\cite{White2009}. The differences between queries at the various stages can provide users with insights in the form of result set changes and awareness of how the query evolves \cite{Krause2016IterativeCohort}. Our system makes use of the notion of \emph{query differences} through difference views of the visual graph query building process. We present a novel interface that displays the changes made to the prototype graph by the user in a single interface, in both the query and result views. Our system, furthermore, utilizes the difference features to incorporate change requests in natural language using a \gls{lm}-based interface. Our novel query interface enables users to instruct the system to add or delete parts of the query graph. This system allows for input in natural language, which is matched to the ontology based on semantic similarity.
This matching enables our system to be robust in the face of incomplete user knowledge about the ontology, and constrains the possible change sets to the limits imposed by the ontology.

\section{Related Works}
\label{sec:rel-works}

Existing graph query builders provide systems to build a single query for a single final information need or question that the user has~\cite{francart2023sparnatural, Vargas2019RDF, Ferre2017SPARKLIS, LIU2022108870, GARCIA2022101235, Clemmer2011, Lissandrini2020}. \emph{SPARNATURAL}~\cite{francart2023sparnatural}, \emph{SPARKLIs}~\cite{Ferre2017SPARKLIS}, and \emph{RDF explorer}~\cite{Vargas2019RDF} are examples of query builders that offer a direct visual interface to build prototype graphs for the query specification, yielding SPARQL queries with limited visual result set exploration. \emph{KGVQL}~\cite{LIU2022108870} relates a visual query language to SPARQL and maps it directly between result sets and the query, or prototype, graph. \emph{Smeagol}~\cite{Clemmer2011} operates similarly, where the user queries the \gls{kg} by example and the system generalizes the query to retrieve more general results. Other approaches, such as \emph{Rhizomer}~\cite{GARCIA2022101235}, utilize multiple views to guide users toward their information needs. Other guidance approaches utilize information retrieval techniques to suggest query expansion options~\cite{Lissandrini2020} and foster \enquote{full} queries. None of these systems, however, consider the query evolution as a relevant aspect of the change in information need or exploratory search~\cite{White2009}, nor do they consider the role of visual representation of the differences. 

Another avenue of query interface systems use \glspl{lm} to provide answers based on natural language queries~\cite{lei2018ontology, Sannigrahi_2024, emonet2024llsparql, meyer2024assessingsparql}. These works, however, provide only singular answers or queries form a natural language query. Our system, however, maps the \gls{lm} output to the visual query interface and uses it to evolve and change the query based on the changing information need of the users.



Showing the differences and similarities between graphs -- i.e., visual graph comparison -- has also been subject to various studies~\cite{Beck2017Taxonomy}. 
For example, there are approaches to illustrate the partial similarities between two graphs by identifying common subgraphs. 
Such can be merged by determining node positions which do not change (much), creating a (super) graph~\cite{diehl_preserving_2001}. 
Alternatively, they can be displayed side-by-side, with a third view showing local similarities in a tripartite multi-view system~\cite{andrews_visual_2009}. 
There are also approaches for weighted graphs, which illustrate the differences in edge weights through a visual encoding of the respective edges~\cite{alper_weighted_2013}. 
Also relevant to our design are approaches for comparing an arbitrary number of (relatively similar) graphs. 
Such can be achieved, e.g., by showing their connectives in a small multiples visualization~\cite{behrisch_visual_2013}, or by stacking the planar graphs in a third dimension, thus giving a 2.5-dimensional view~\cite{brandes_visual_2004}.  Another approach, \emph{EditLens}, utilizes a lens system to show changes performed on a larger graph~\cite{Gladisch2014}. 
As we are dealing with progressive graphs, we rely on highlighting through color to indicate changes. Our system, furthermore, operates on smaller sub-graphs and within a novel domain, query graphs, and applies the concept to exploratory search.


\section{Methodology}
\label{sec:meth}

\subsection{Graph Definition}

Our \gls{onset} system builds upon the notion of a prototypical graph $G_p \coloneqq (N_p, E_p)$ that serves as the blueprint for all retrieved instances. This graph representation is similar to \gls{bgp}~\cite{hogan2021knowledge}, a system to express simple graph patterns for queries over \glspl{kg}. Our definition extends this by adhering to the schema imposed by the ontology. The sub-graph $G_{proto}$ of the ontology consists of the nodes $N_{proto}$, each one an instance of the classes $\mathcal{C}$, and edges $E_{proto}$, each one a link of link types $\mathcal{L}$. The multiset of nodes can contain a class 
multiple times, and a link can only span the allowed end and start types (within the class hierarchy), i.e.
\begin{align*}
    N_{\textit{proto}} & \coloneqq \big\{n_1, \ldots, n_m\big\} \subseteq \mathcal{C} \,,                                           \\
    E_{\textit{proto}} & \subseteq \big\{e_i=(n_t, l_j, n_h) \;\big|\; \mathrm{subtypeof}(n_t, \,\mathrm{fromtype}(l_j)),           \\
                       & \qquad \mathrm{subtypeof}(n_h,\,\mathrm{totype}(l_j))\big\} \subseteq (N_p\times \mathcal{L}\times N_p)\,.
\end{align*}
Each node can be extended using additional sub-queries related to the node's classes properties $p\in\mathcal{P}_{N_{\textit{proto}}}$, which can be either constraints $\text{constraint}(p, \text{cond.})$ or property values $\text{value}(p)$
\begin{align*}
    S_{N_{\textit{proto}}}\coloneqq \big\{s_1, \ldots\big\} \subseteq  \text{constraint}(\mathcal{P}_{N_{\textit{proto}}}, \text{cond.}) \cup   \text{value}(\mathcal{P}_{N_{\textit{proto}}}, \text{cond.})
\end{align*}
The resulting instance graphs are queried from the database by constructing the SPARQL from the class, link, and sub-query constraint sets and retrieving the results.

\subsection{Difference Graphs}
To construct the difference between graphs, we take two graphs, $G_{proto, l}$ and $G_{proto, r}$, and construct the set differences, which will yield the additions and removals of nodes. We compare the individual nodes and constraints by serial numerical identifiers, while the result graphs are compared using their connected instance nodes.

The added and deleted sets are therefore constructed by comparing the identifiers, i.e.
\begin{align*}
    N_{\textit{proto}, \textit{add}} \coloneqq    & \big\{ n_{\textit{proto},r} \in  N_{\textit{proto},r} \vert                                                                          \\
                                                  & \text{id} (n_{\textit{proto},r}) \notin \{ \text{id}(n_{\textit{proto},l}), n_{\textit{proto},l}\in  N_{\textit{proto},l}\} \big\}   \\
    N_{\textit{proto}, \textit{del}} \coloneqq    & \big\{ n_{proto,l} \in  N_{proto,l} \vert                                                                                          & \\
                                                  & \text{id} (n_{\textit{proto},l}) \notin \{ \text{id}(n_{\textit{proto},r}),n_{\textit{proto},r} \in  N_{\textit{proto},r}\} \big\}   \\
    N_{\textit{proto}, \textit{shared}} \coloneqq & \big\{ n_{\textit{proto},l} \in  N_{\textit{proto},l} \vert                                                                        & \\
                                                  & \text{id} (n_{\textit{proto},l}) \in \{ \text{id}(n_{\textit{proto},r}),n_{\textit{proto},r} \in  N_{proto,r}\} \big\}.
\end{align*}
Similarly, the added and deleted sets of the sub-queries can be constructed using similar operations performed on the unchanged nodes $ N_{proto,shared}$, i.e.
\begin{align*}
    S_{N_{\textit{proto},\textit{shared}}, add} \coloneqq & \big\{ s_{\textit{proto},r} \in  S_{N_{\textit{proto},shared},r} \vert                                                           \\
                                                          & \text{id} (s_{\textit{proto},r}) \notin \{ \text{id}(s_{proto,l}), s_{proto,l}\in  S_{N_{proto,l}}\} \big\}                      \\
    S_{N_{\textit{proto},\textit{shared}}, del} \coloneqq & \big\{ s_{\textit{proto},l} \in  S_{N_{\textit{proto},\textit{shared}}} \vert                                                  & \\
                                                          & \text{id} (s_{\textit{proto},l}) \notin \{ \text{id}(s_{\textit{proto},r}),s_{\textit{proto},r} \in  S_{N_{proto,r}}\} \big\}.
\end{align*}

We, however, enable the user to change the sub-queries, e.g. they can alter the parameters or operations performed within these. To model these changes, we define a third set that contains all changed sub-queries, determined by a $\text{changed}(s)$-operator, i.e.
\begin{align*}
    S_{N_{\textit{proto},\textit{shared}}, chg} \coloneqq & \big\{ s_{\textit{proto},s} \in  S_{N_{\textit{proto},\textit{shared}}} \vert & \text{is changed} (s_{\textit{proto},s})\big\}.
\end{align*}
These difference sets are then displayed directly within the query view if the user is tracking their changes, and updated dynamically.  We highlight the added nodes in green, and the deleted ones in red\footnote{We offer a colorblind mode to aid users with Red-Green color blindness.}

\subsection{Automatic Result Set Overview}
To enable a more in-depth result overview, we provide value-wise fetching of node attributes using the sub-query elements mentioned above. The user may be searching for the birthdates of people participating in the Olympics in a specific year and wants to know the distribution of these dates, as there might be hundreds of participants. The result set overview enables a quick visualization of the attributes of the nodes within the graph using standard techniques for common data types. We employ a heuristic to select the most suitable visualization type based on the current setting. If the user chooses only one value, we show the feature distribution. If the feature is continuous we group it into a set number of buckets, if it is discrete we show the most common discrete types.
Furthermore, if the user chooses two value attributes, we either display a heatmap if there are too many data points or use a scatter plot to show the distribution of the results. For the Olympiad's example, the system would choose the histogram plot. We integrate these result set charts with our difference view by comparing the different result sets of the two prototype graphs, as long as the values selected to be retrieved still match. These could be the birthdate distributions of a specific country compared to the overall distribution for our example. We also allow users to download difference sets to apply their own statistical and visual evaluations outside our system, provided they are satisfied with the information from our queries.

\subsection{Integration of a Natural Language Query System}

We utilize the difference views to manage the integration of a \gls{lm}-based query feature, allowing users to view the possible changes the \gls{lm} would make based on a natural language query before accepting or rejecting them. This is especially helpful if the \gls{lm} makes a mistake by suggesting the wrong links or hallucination \cite{Ji2023}. Other failure cases include an information need by the user that the \gls{kg} cannot satisfy, but the user specified in their query. The assistance system may still return some changes that are loosely related to the user request, but the user can still modify them before accepting the changes.

The query parsing to a structured change set follows a three-step architecture to achieve robust query modifications for fuzzy user input:
\vspace*{-0.75em}
\begin{enumerate}
    \itemsep-0.2em
    \item The query is passed to a \gls{lm}, generating embeddings of the query context and used to retrieve $k$ semantically similar classes and links that could be the intended true classes and links. These are then used to construct a highly constrained schema that only allows the connections specified by the ontology.
    \item This highly constrained schema is then used for the output constraints of the \gls{lm} with strict rules on the allowed class and link types.
    \item Finally, the output is corrected as the schema itself does not fully enforce a consistent graph change, which can be corrected using a few rules. These include adding missing nodes or flipping directions of links if they are reversed w.r.t to the ontology.
\end{enumerate}
\vspace*{-0.75em}

This change set is then applied to the prototype graph $G_{proto}$, and the difference view is automatically activated, including a first initial difference calculation according to the schema outlined above.

\begin{figure}[t]
    \centering
    \includegraphics[width=1\linewidth]{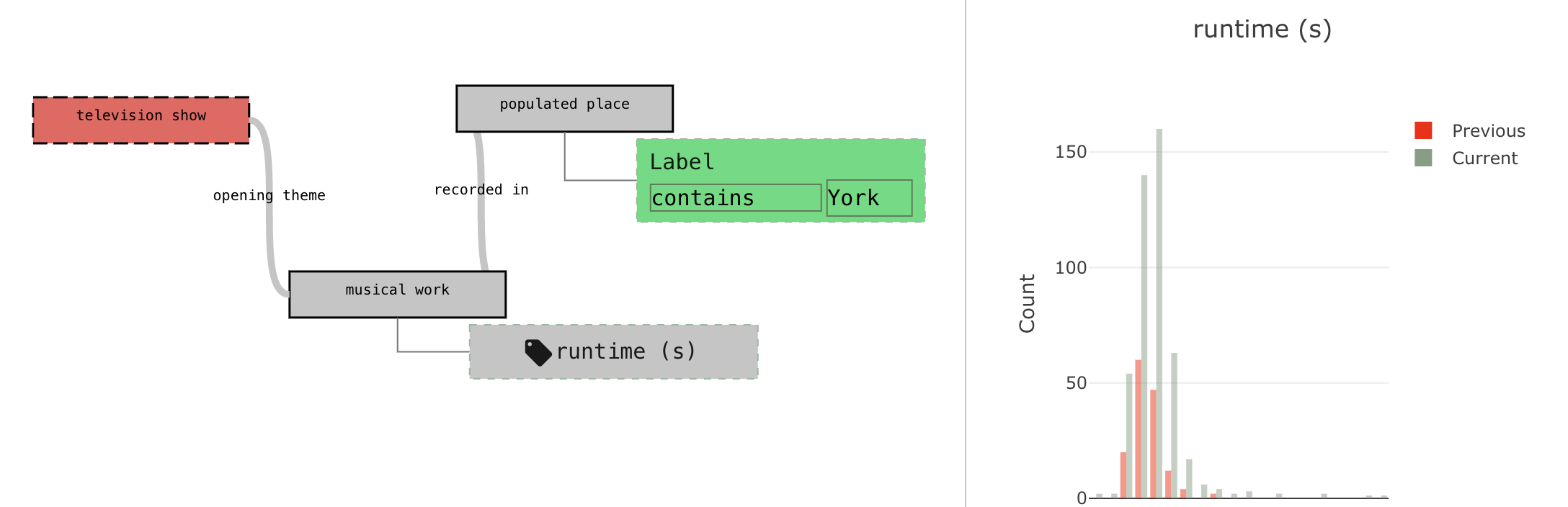}
    \caption{Simple difference query comparing the distribution between two different queries related to musical works.}
    \label{fig:dbpedia:single_val}
\end{figure}

\section{Implementation}
\label{sec:impl}

The outlined concepts are integrated into our \gls{onset} system, focusing on fast user responses even on larger ontologies. We also base our entire stack on open-source systems, allowing institutions or users to start and customize their systems.

To achieve fast user responses, even on more complex queries and on commodity hardware, we use qlever~\cite{Bast2017}. This SPARQL query engine outperforms most existing engines in both speed and system requirements. This speedup enables our system to serve and display updates as the user builds their query, aiding the user in retrieving non-empty sets and showing intermediate results to guide the search.

To maintain our intended fast user responses, we only compute the embeddings of links and classes during the first startup with a specific ontology. We store our resulting embeddings in PostgreSQL\footnote{\url{https://www.postgresql.org/}} with the help of the pgvector extension \footnote{\url{https://github.com/pgvector/pgvector}}, allowing fast retrieval given a query embedding. To generate these embeddings quickly, even on commodity hardware, we use the \verb|stella_en_400M_v5|\footnote{\url{https://huggingface.co/NovaSearch/stella_en_400M_v5}} model, which is, at the time of submission, the best-performing smaller model w.r.t. the massive text embedding benchmark~\cite{muennighoff2022mteb}. We utilize the Hermes Llama 3.2 8B model for our structured output generation in the natural language interface~\cite{teknium2024hermes3technicalreport}.

Our user interface builds on Vue.js\footnote{\url{https://vuejs.org/}}, in combination with three.js\footnote{\url{https://threejs.org/}} and D3.js~\cite{2011-d3}. All the used database systems and models are open-source and open-weight, providing state-of-the-art performance in their respective fields while still being able to run on commodity hardware. We provide the code to our system on \url{https://github.com/Dakantz/OnSET}.
\section{Results  \& Case Study}
\label{sec:results}

\begin{figure}[t]
    \centering
    \includegraphics[width=1\linewidth]{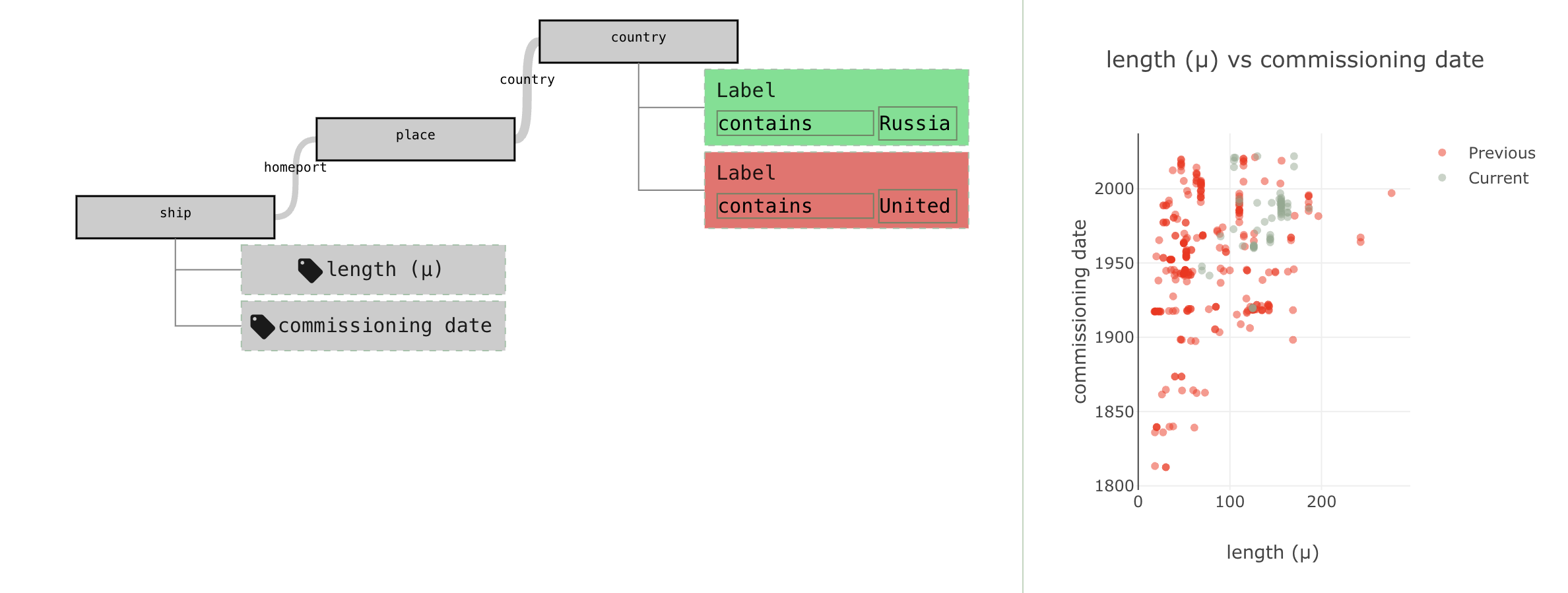}
    \caption{Difference of two variable distributions (length and commissioning date) of ships between different countries.}
    \label{fig:dbpedia:two_val}
\end{figure}

\subsection{DBpedia}

DBpedia is a comprehensive \gls{kg} that utilizes Wikipedia data to construct a rich ontology comprising 768 classes, linked to approximately 4,233,000 instances \cite{Lehmann2015DBpediaA}. We demonstrate how our system can facilitate exploratory search of this \gls{kg} to discover interesting facts and assist users in building and extending their queries using natural language.

\subsubsection{Exploratory Search}

Our first use case for an exploratory application highlights how the difference graphs enable users to compare and extract information between different queries easily.

In our first example in \cref{fig:dbpedia:single_val}, the user first searches for opening themes of television shows that are recorded in any popular place. They select the runtime in seconds to be plotted. The users, however, decide to contrast this initial information without the opening theme constraint and add the constraint "York" to the label of the populated place. Our interface displays this shift, first, by marking the added nodes in green and the deleted parts in red, with additional indications in the borders of the nodes.  The second change is the difference view in the result distribution, showing how the runtime compares between the queries. The users could observe that there are many more results for the second query, and that the opening themes have, on average, a shorter runtime. 
Our second example in \cref{fig:dbpedia:two_val} illustrates how a user could search for relations of ships within the \gls{kg} of DBpedia. This query revolves around the relationship between the length of a ship, its commissioning date, and the country where the ship's homeport is located. To apply our difference views, the user first searches for all ships within the United States. These ships are then set in contrast to those from Russia, immediately illustrating the difference in distribution to the user. They could observe that the United States had had many more ships much earlier. It is also clear that the size and number of ships have increased rapidly over the years, with Russian ships being significantly less prominent.

\begin{figure}[t]
    \centering
    \includegraphics[width=1\linewidth]{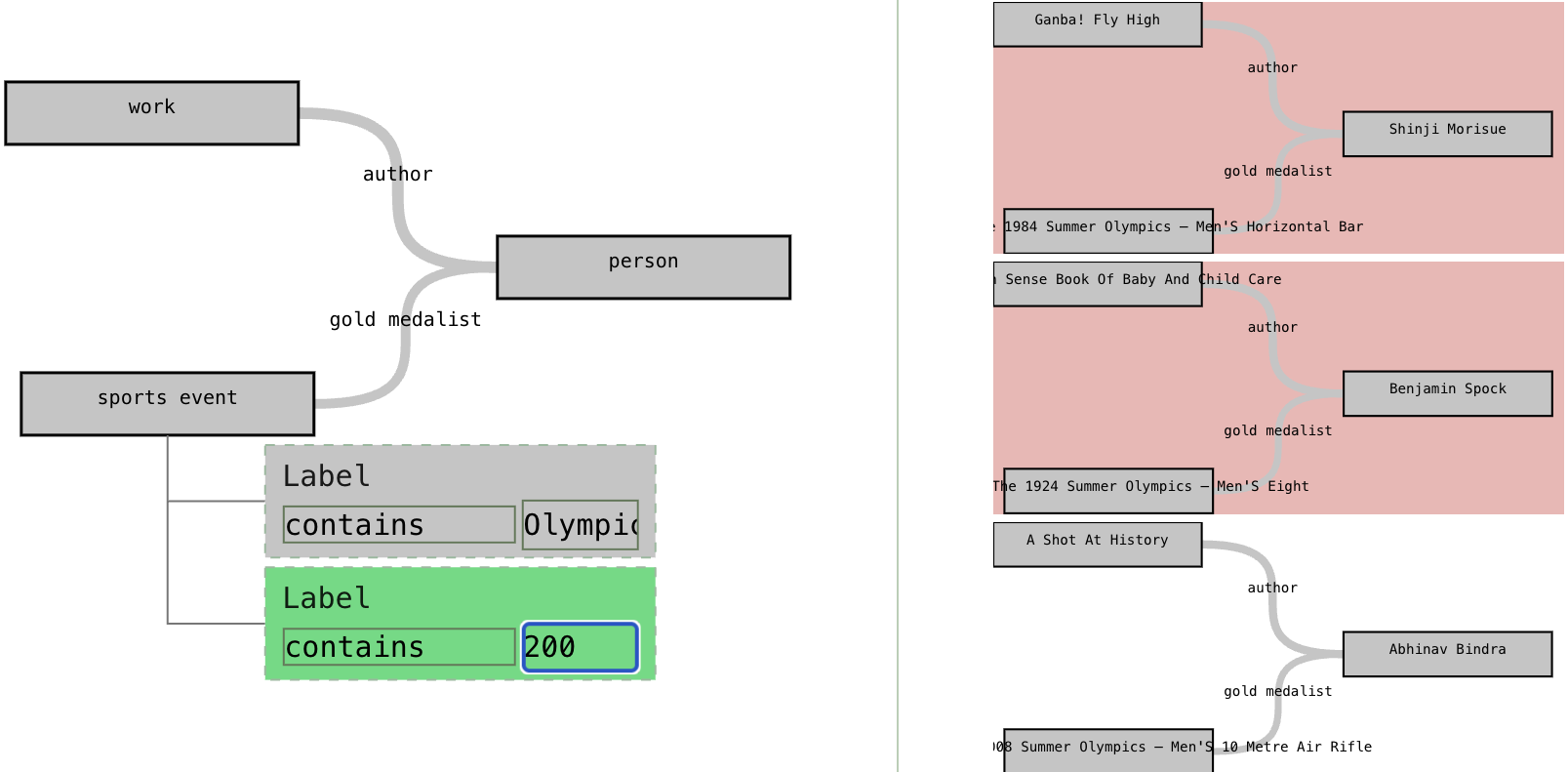}
    \caption{Differences in instance results for Olympic sports event with an additional filter.}
    \label{fig:dbpedia:instance_diff}
\end{figure}

Our third example considers smaller result sets, where the query might be more specific and the result relations are of interest. For this use, we provide different views in the instance result view, showing individual instances of the result graph to the user. We again calculate the difference across all results and highlight added and removed instances. In this case, the user queries for persons who are both authors of a work and gold medalists in a sports event, as shown in \cref{fig:dbpedia:instance_diff}. There are, naturally, only a few instances of this particular relation within the graph, and the user may be interested in the specific works by Olympians who participated in the games of the 2000s, adding a label constraint. We show the user the difference in the result set, emphasizing how the result set might have reduced in size due to the additional constraint.

\begin{figure}[t]
    \centering
    \includegraphics[width=1\linewidth]{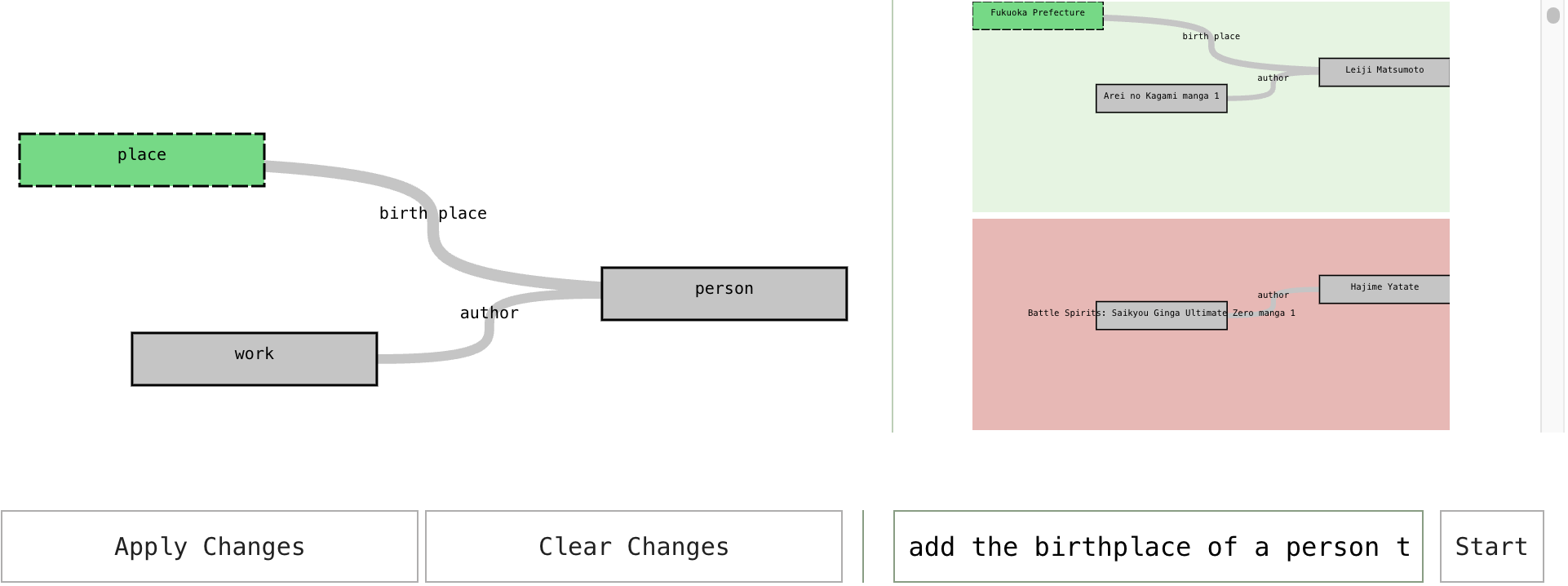}
    \caption{\gls{lm} assistance to edit the query using natural language. The user prompts the \gls{lm} to \enquote{add the birthplace of a person} and is presented with the suggested change set and direct effect in the result instances.}
    \label{fig:dbpedia:lm_assist}
\end{figure}

All three examples illustrate how our query difference graph can immediately convey the relevant differences between different queries, fostering an understanding of result views that are relevant to the user's evolving interests during exploration, and relating the results to prior information needs. 

\subsubsection{Natural Language Interface}

We also demonstrate our natural language interface and how it can be used to extend the query graph. We allow the user to enter instructions to modify the query. The resulting output of the \gls{lm} is constrained to the ontology, specifically DBpedia. In this example, as shown in \cref{fig:dbpedia:lm_assist}, the user requests that the place of birth should be added to the query graph, and the \gls{lm} performs the modification and triggers our difference view directly. This enables the user to, on the one hand, view the modifications made during this step, and, on the other hand, revert any potential mistakes with just one click and reformulate their prompt.

\begin{figure}[t]
    \centering
    \includegraphics[width=1\linewidth]{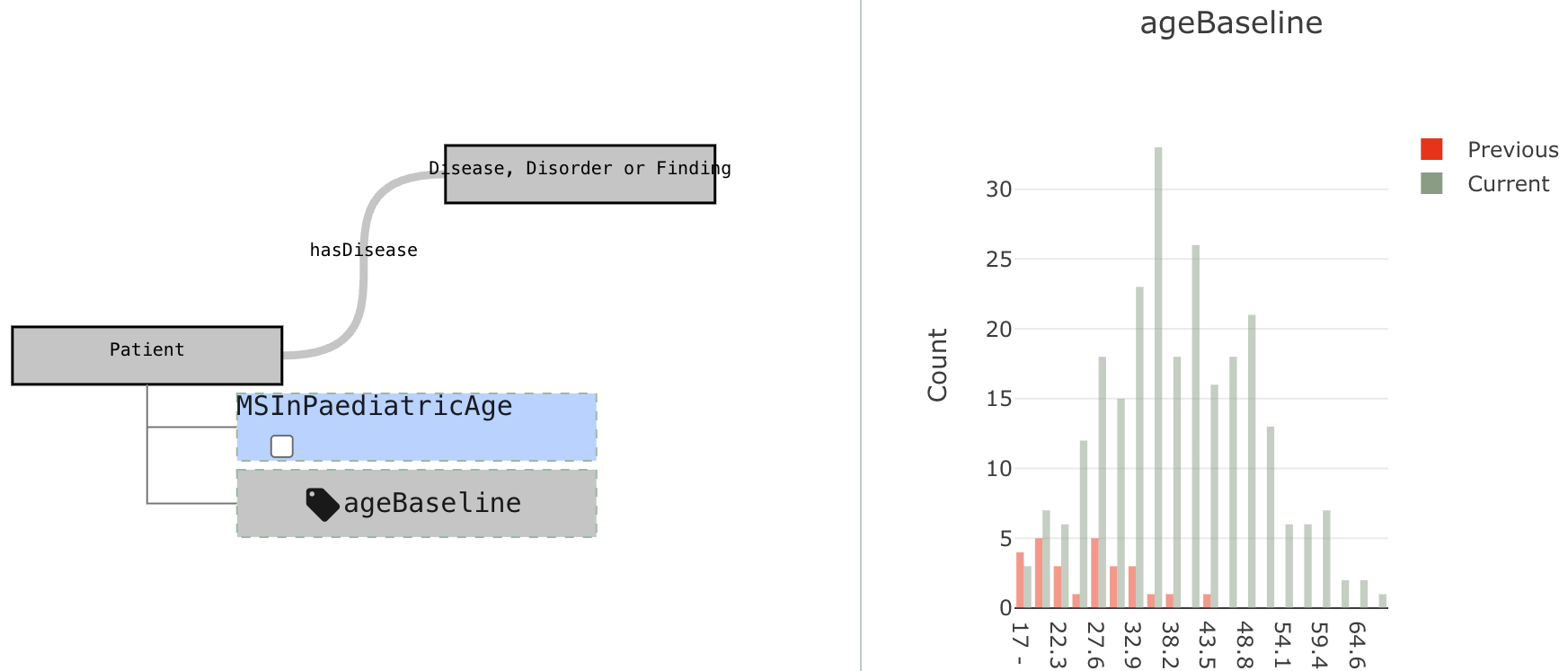}
    \caption{The \gls{bto} \gls{kg} queried using the difference view to explore the relationship between different semantic attributes.}
    \label{fig:bto:age}
\end{figure}

\subsection{\gls{bto}}

Our second explored ontology is the \gls{bto} \cite{Faggioli2024}, which maps \gls{als} and \gls{ms} data from clinical trials to a \gls{kg} using anonymized data. This medical ontology serves as a basis for applications beyond exploratory use cases.

For this case study in \cref{fig:bto:age}, we use just a small query graph to keep our study simple. We assume that the user is interested in the age distribution of patients suffering any disease or condition and filters on the property \enquote{MSInPeaediatricAge}. The user then changes the filter to the negation while the difference view is enabled, which triggers the difference view for the histogram plot. While these two variables are semantically related, a secondary effect appears to be present in the data, as the distribution does not simply cut off, but changes completely. To investigate this effect further, the user can switch to the instance result view and examine individual results, or export the results from our tool and conduct statistical analyses using their preferred tool.

\section{Discussion \& Conclusion}
\label{sec:disc}


This paper presents a novel difference view for graph query building that allows users to explore the evolution of their graph query in an integrated view with the subgraph, constraints, and result sets. We also integrate a \gls{lm} interface to facilitate exploration with little prior knowledge of the graph and the user interface, leveraging semantic retrieval and fuzzy matching to the strict ontology structure. We allow, through the difference implementation, reversal of the changes to reduce the effect of invalid modifications by the \gls{lm}, and more experimental searches by the users.

The resulting tools enable non-expert users to explore relations of interest within the \gls{kg}, including the dependencies of attributes within the queried sub-graph. Our difference views, specifically, foster the understanding of differences of sub-graphs and their attributes and distributions over the whole result set. This effect in distributional change is especially evident for the change in attribute filters, which we demonstrate through several case studies on different ontologies and  \glspl{kg}, the DBpedia and \gls{bto}. While the DBpedia use cases demonstrate how our system informs users in exploratory use cases, the \gls{bto} illustrates the applicability of our tool to experts in their respective fields. They can build example graphs and query for distributional changes in medical disease attributes, exploring these differences within our tool.


\subsection{Future Work}
\label{sec:future-work}

While our difference views encompass a highly expressive graph structure, complete with constraints, attributes, and links throughout the graph, we still see areas of improvement for the integration of the visualization technique within \gls{onset}. One of the steps towards a complete evolutionary representation could be the versioning of the graph over multiple iterations, where the visual representation of small steps and grouping changes automatically poses a challenge. Other improvements towards a guided exploration system could include link and change recommendations based on prior interests, and fine-tuning the \gls{lm} to enhance the suggested changes as we utilize few-shot prompting to refine the suggestions currently.



%
%

\bibliographystyle{abbrv-doi}

\bibliography{references}
\end{document}